\RequirePackage{fix-cm}
\documentclass[smallextended]{svjour3}       
\smartqed  
\usepackage{graphicx}
\usepackage{epstopdf}
\usepackage{mathptmx}      
\usepackage{latexsym}
\begin{document}

\title{Overview of ISM bands and Software-defined Radio Experimentation}


\author{Abhaykumar Kumbhar}


\institute{Department of Electrical and Computer Engineering, Florida International University\at
              Miami, FL 33174 \\
              \email{akumb004@fiu.edu}
}

\date{Received: date / Accepted: date}

\maketitle

\begin{abstract}

Wireless systems using low-power wireless communication protocol are rapidly gain popularity in the license-free industrial scientific, and  medical (ISM) frequency bands. One such emerging trend in ISM frequency bands is home automation. Historically, all the home devices were once unconnected, today are now being connected either by a wired or wireless connection. The low-power wireless communication protocols enable integration of all the digital home devices into a single system and enhance end user product experience. The rapid prototyping of these end user product using programmable radio system such as software-defined radio (SDR) has been the game-changer in the field of home automation. In this article, we present an overview of popular ISM bands in different regions of the world. Furthermore, we analyze the signals from the wireless electric switch using a software-defined radio in 433 MHz ISM band.  

 
\end{abstract}

\section{Introduction}
\label{intro}
The ISM frequency bands are scattered from 6. 78 KHz to 245 GHz in the radio spectrum, has been reserved for  industrial, scientific and medical purposes. This band was originally designed for used applications such as microwave, medical equipment, process heating, and types of electrodeless lamps. However, over the years, wireless communication standards and equipment have been developed and manufactured that are capable of operating in these bands without causing any interference to existing ISM device operations. These non-ISM devices are usually short-range, low power communications devices and systems such as cordless phones, Bluetooth devices, near field communication (NFC) devices, RFID, ZigBee, and wireless computer networks. Depending on the international telecommunication union (ITU) region, every region has its own regulation for the operation of these non-ISM devices. These rules specify the maximum transmission power, duty cycle,  and signal periodicity.

Wireless home automation networks have found its footprints in the ISM bands as the licensed/unlicensed systems. A home automation system can be outlined as home's electrical devices are connected to a central system  and are automated based on user inputs \cite{alkar2005internet}. For instance, this centralized system enables a user to control and monitor lighting, heating, ventilation, air conditioning, appliances, security locks, and other related systems. Thus, providing an improved ease of handling, convenience, efficiency, comfort, and utmost control. This centralized control and monitor system can be either computer, or tablet, or smart-phone. A wireless home automation network consists of wireless embedded sensors and actuators that enable monitoring and remote control of wireless devices \cite{gomez2010wireless}. For instance, at a mere push of the button an electric bulb can turn on/off remotely \cite{mainetti2011evolution}.

Historically, SDR were expensive and usage were restricted only to mobile base stations. With recent development in the low-cost SDR platform, it successfully led to mass production of SDR-based terminal. The lower cost of SDR hardware and availability of open source software, has opened several portal with new applications at low cost. These SDR application opportunities can also be extended to current next big i.e., home automation and it has been experiencing rapid growth \cite{SDR0}.
 
\subsection{Contributions} 
In this paper, we carry out a brief study of various ISM frequency bands and discuss challenges surrounding these frequency bands. In the Table. \ref{433Comparetable}, presents a comparative overview of the popular ISM bands i.e., 433 MHz, 868 MHz, 915 MHz, and 2.4 GHz. This tables we discusses the pros, cons, and frequency range. Furthermore, in a SDR experiment we study ISM device control signals captured from Eteckcity device. The paper is outlined with Section~\ref{ISMband} presenting a brief overview of existing ISM bands. SDR has been enabling new technologies and Section~\ref{SectionHAI} of the article gives introduction to SDR. Furthermore, Section~\ref{SectionSDRExp} furnishes the experimental study using SDR. Section V communicates the future goal of this article, and Section VI concludes the paper.

\section{ISM Bands}
\label{ISMband}
The new generation of cost-effective wireless technology has been driven by relatively lower hardware and software cost and low power consumption. Furthermore, has also led to the rise of new types of networks with wide range of applications for security, home and building automation, smart meters, and personal medical devices. However, these wireless technologies have relatively low data rate and operate within a closed wireless network. These wireless technologies operate within wide range of frequency bands known as the ISM radio band as show in \cite{433_0}. Within ISM bands, the 2.45 GHz, 915 MHz,  868 MHz, and 433.92 MHz are most widely used bands in the field of IoT and home automation. Other ISM bands include 6.78 KHz, 13.56 KHz, 27.120 KHz, 40.62 MHz, 5.8 GHz, 24.125  GHz, 61.25 GHz, 122.5 GHz, and 245 GHz \cite{ituIsmProc}.

The 433 MHz wireless communication is popular amongst the IoT solution makers and hobbyists. This frequency band is intended for applications where transmission stops within five seconds such as switches, temperature sensors, wall sockets, and window/door contact sensors \cite{433_1}. The 433 MHz is the most widely used ISM band through Europe and most of rest of the world.  An overview of 433 MHz pros and cons is presented in Table \ref{433Comparetable}. The 433 MHz band has a frequency range of 433.050 MHz to 434.790 MHz, which includes 69 channels with channel spacing of 25 kHz. Furthermore, the standard characteristics of devices operating in 433 MHz is also shown in Table \ref{433DeviceAttributes}.

The frequency range of 902 MHz - 928 MHz with 915 MHz center frequency is available in ITU region 2 with some exceptions. Whereas, 863 MHz - 870 MHz frequency range is an ISM band in Europe only. In the United States, the Federal Communications Commission (FCC) regulates the 915 MHz ISM band. This ISM band doesn't place any restrictions on the type of application or duty cycle. The whole 868 MHz and 915 Mhz band are considered as a tunning range. Only parts of this tuning range are operationally available in some countries \cite{SDR}. Typical 915 MHz application include SCADA and RFID. Depending on the device radio system and they can support ASK, FHSS, FSK, and GFSK type of modulation scheme. The devices operating in these bands can have transceiver voltage of upto 6 V and the RF power is restricted to 1 mW in the United States and 25 mW in Europe. In the United States the device have no limitations on the duty cycle, however in Europe the duty cycle is restricted to $1\%$ \cite{shortRange2}.

The global availability of 2.4 GHz band has made it one of the popular ISM bands, as it is suitable for low-cost radio solutions such as wireless personal area network and wireless local area network (WLAN). Bluetooth, WLAN 802.11b, and ZigBee are some of the applications supported in 2.4 GHz. Due to heavy traffic in the 2.4 GHz band, other ISM bands in lower frequencies of 915 MHz,  868 MHz, and 433 MHz have been considered in ITU region 1 and ITU region 2 \cite{2.4Ghz}. The number of channels in 2.4 GHz is variable and depends on the wireless technology operating in the band. For instance, WLAN has 13 channels with a static channel width of 22 MHz. Whereas Bluetooth has 79 channels and dynamically allocated channel 
bandwidth of 15 MHz. Similarly for ZigBee, the number of channels are 16 and has a static channel width of 3 MHz \cite{2.4Channels}. Depending on the device radio system and they can support AFH, DSSS, GFSK, and FHSS type of modulation scheme. The devices have minimum transceiver voltage of 2 V and maximum permissible output power of 1 W or 30 dBm.

\begin{table}[!htbp]
\centering
\caption{ Comparison of various ISM bands \cite{433_0}. where $f_c$ is center frequency.}
\label{433Comparetable}
\begin{tabular}{p{0.75cm}p{2cm}p{2cm}p{1.75cm}p{1cm}p{1.075cm}p{1cm}} 
\hline\noalign{\smallskip}
Band  & Pros & Cons & Frequency range & $f_c$ & ITU reg. & References\\ 
\noalign{\smallskip}\hline\noalign{\smallskip}
433 MHz  
& 1)Longer range. 2) Better indoor signal penetration. 3) Low path loss. 
& 1) Overcrowded in some frequencies. 2) Relatively low quality and expensive radio system. 
& 433.050 - 434.790 MHz. 
& 433.92 MHz.
& Reg. 1 (Europe).
& \cite{433_0,433_1,2.4Ghz,SDR,shortRange}\\

868 MHz 
& 1) Reasonable range. 2) Relatively less congested band.
& 1) Complex band plan.
& 863 - 870 MHz.
& 868 MHz
& Reg. 1 (Europe).
& \cite{2.4Ghz,SDR,shortRange,shortRange1}\\

915 MHz
& 1) Large range upto 1 km. 2) Smaller antenna sizes.
& 1) Limited applications. 2) Needs better regulations for ISM usage. 3) High power consumption of devices.
& 902 - 928 MHz.
& 915 MHz.
& Reg. 3 (AUS/NZ), Reg. 2 (USA/CAN).
& \cite{2.4Ghz,SDR,shortRange}\\

2.4 GHz  & 
1) High data-rate. 2) Low cost versatile radio systems.
& 1) Low range but can be increased using repeaters. 2)Higher path loss.
& 24 - 24.25 GHz.
& 24.125 GHz.
& Reg. 1,2, and 3.
& \cite{2.4Ghz,2.4Channels,SDR,shortRange}\\

\noalign{\smallskip}\hline
\end{tabular}
\end{table}

\subsection{Major Challenges}
Even with several technical advancements, there are major challenges that can hinder efficient operation of an device in ISM band.
\begin{enumerate}
  \item \emph{Harmonization of ISM bands}: As seen in Table. \ref{433Comparetable}, ITU region 1, 2, and 3 have different operating ISM bands. The ISM applications are deployed in designated bands as well as undesignated band. The ISM frequency range varies for global or regional usage and therefore needs harmonization \cite{hormony0}. However establishing radio regulation for ISM band is an ongoing challenge for on a worldwide basis.
  \item \emph{Increasing spectrum demand}: With increasing number of devices operating in ISM bands has resulted in increasing spectrum demand. However with limited availability of spectrum makes the increasing spectrum demand challenging. Nonetheless, solutions like cognitive spectrum approach and operation in unlicensed spectrum can help to met the increasing the spectrum demands. Thus making use of existing spectrum more efficiently and economically.
  \item \emph{Interference coordination}: The harmonious coexistence of ISM devices in a heterogeneous environment with similar technical specification, need to address the potential interference issue. By applying band specific mitigation technique, can facilitate harmonious operation of various heterogeneous ISM devices \cite{hormony0,baccour2013external,moussavinik2013narrowband}.  
\end{enumerate}

\section{Software-defined Radio}
\label{SectionHAI}
The Wireless Innovation Forum, established in the year 1996, is dedicated to endorsing the innovative use of spectrum and advancing radio technologies that support critical communications worldwide such as SDR, cognitive radio and dynamic spectrum access technologies. The main goal of SDR is to provide complementary support to current and future mobile communications by providing configurable radio platform \cite{nesimoglu2010review,srilatha2013knowledge}. 

The software programmability of SDR allows to change the radio's fundamental characteristics such as modulation types, operating frequencies, bandwidths, multiple access schemes, source and channel coding/decoding methods, frequency spreading/despreading techniques, and encryption/decryption algorithms \cite{SDR}. The programming flexibility and cost efficiency has given the product developers and researchers to drive SDR forward in the forming of end products with  wide-reaching benefits. The wireless communication in the home automation provides the necessary infrastructure for bi-directional communication for data collection and delivering control messages. The platform based on SDR technology can provide the necessary communications bridge as seen in \cite{lin2013wireless}. SDR has been experiencing rapid growth due to its user-ready, programmable radio systems, low-cost, and sufficient computational power abilities.  

The authors in article \cite{ulversoy2010software,tuttlebee1999software}, discuss the importance, evolution, and various applications of SDR. With the increasing availability of software tools and platforms for SDR, its potential in the field of radio communication industry is on a steady increase. Furthermore, the role of SDR in home automation and wireless sensors in discussed in \cite{ferrari2011new}. SDR isolates the need for expensive hardware and presents with an ability to adapt to external conditions using software programming \cite{ferrari2011new}.

\section{SDR Experimentation}
\label{SectionSDRExp}
The radio communication system hardware such as amplifiers, mixers, filters, modulators/demodulators, and detectors are implemented using the software in a SDR \cite{sdr1,sdr2,sdr3}. By means of a variable-frequency oscillator, mixer, and filter, SDR receiver can tune into desired frequency or a baseband \cite{sdr1}. The low-cost infrastructure provided by the SDR makes it an integral part of researchers, hobbyist, and rapid prototyping. The goal of this section is to carry out the basic experimentation using SDR receiver HackRF \cite{hackrf} and capture the signals from wireless remote for the electric switch.

\subsection{SDR}
\label{whatIsSDR}
SDR is a radio communication system where components have been typically implemented by means of software on a personal computer or a embedded system. The rapidly evolving capabilities of digital electronics and software, has rendered the concept of SDR many practical processes which used to be only theoretically possible. The basic block diagram of SDR is shown in \cite{sdr2}. Software radio has its application from real-time military applications to cellular applications and experimentation in academic field to rapid prototyping of home-automation. 

\begin{figure} [!htbp]
\centering
\includegraphics[width=4cm,height=4cm,keepaspectratio]{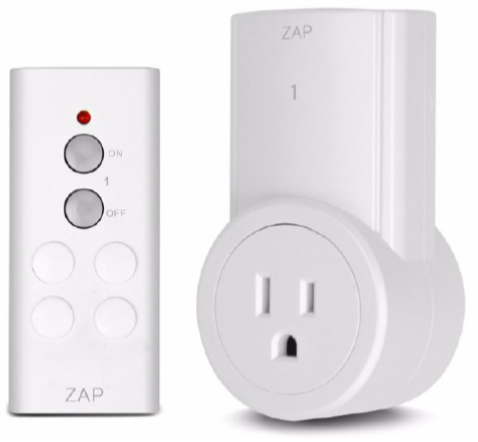}
\caption{Etekcity wireless controlled electric switch socket outlet with remote. This 433 MHz ISM device has 100ft range and works through doors, floors, and walls \cite{Eteckcity_Device}.}
\label{fig:EtekcityDevice}
\end{figure}

\begin{table}[!htbp]
\centering
\caption{ 433 MHz band device attributes.}
\label{433DeviceAttributes}
\begin{tabular}{lp{8.5cm}} 
\hline\noalign{\smallskip}
Attribute & Description \\ 
\noalign{\smallskip}\hline\noalign{\smallskip}
Antenna &  Quarter-wave length antenna i.e., approximately 17.2 cm\\
Modulation & Depends on the device radio system and can support 2FSK, GFSK, 4FSK, ASK, OOK and MSK modulation scheme\\
Transceiver voltage & Upto 12 V\\
RF Power &  Maximum permissible output power 10 mW\\              
Type of use & Intermittent control signals and periodic transmissions\\
Duty cycle & upto 100 \%\\
\noalign{\smallskip}\hline
\end{tabular}
\end{table}

\subsection{Setup}
\label{subsectionSetup}
\begin{figure} [!htbp]
\centering
\includegraphics[width=9cm,height=3.5cm,keepaspectratio]{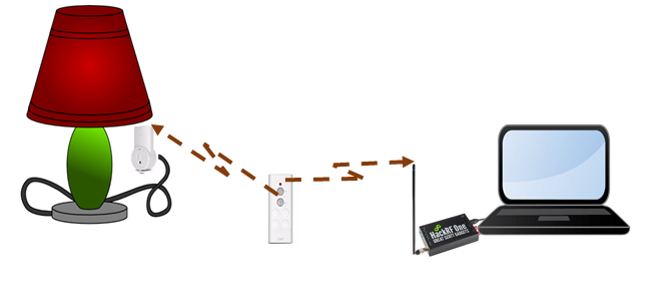}
\caption{SDR experimental setup.}
\label{fig:expSetup}
\end{figure}

\begin{figure} [!htbp]
\centering
\includegraphics[width=12cm,height=3cm,keepaspectratio]{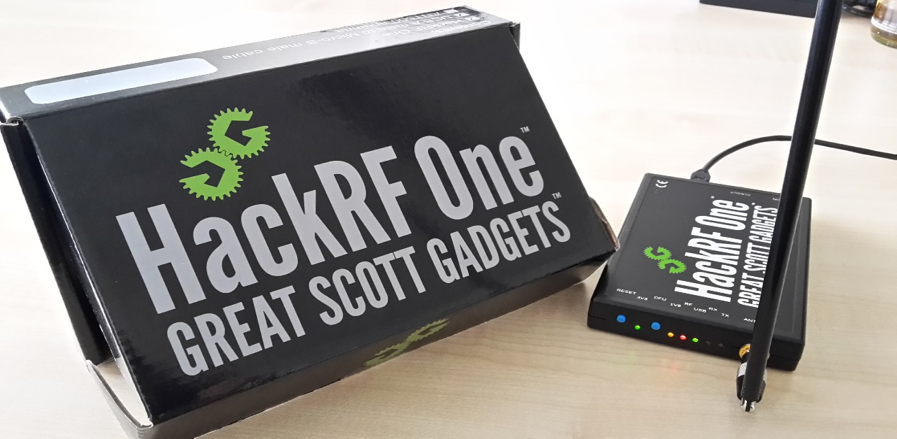}
\caption{HackRF device used in the experimentation.}
\label{fig:hackRf}
\end{figure}

In the current experimental setup, the electric lamp is connected to the electric outlet via etekcity remote controlled wireless switch shown in Fig. \ref{fig:EtekcityDevice} and Fig. \ref{fig:expSetup}. HackRF SDR devie shown in Fig. \ref{fig:hackRf} is used to sniff the control packets wirelessly. HackRF is a cheap SDR transceiver, since the chip allows transferring the raw I/Q samples to the host, which is officially used for DAB/DAB+/FM applications. Table. \ref{433DeviceAttributes}, reveals the attributes of device operating in 433 MHz ISM band.

\subsection{Initial signal evaluation using SDR\#}
\label{initSigVal}
\begin{figure} [!htbp]
\centering
\includegraphics[width=8cm,height=8cm,keepaspectratio]{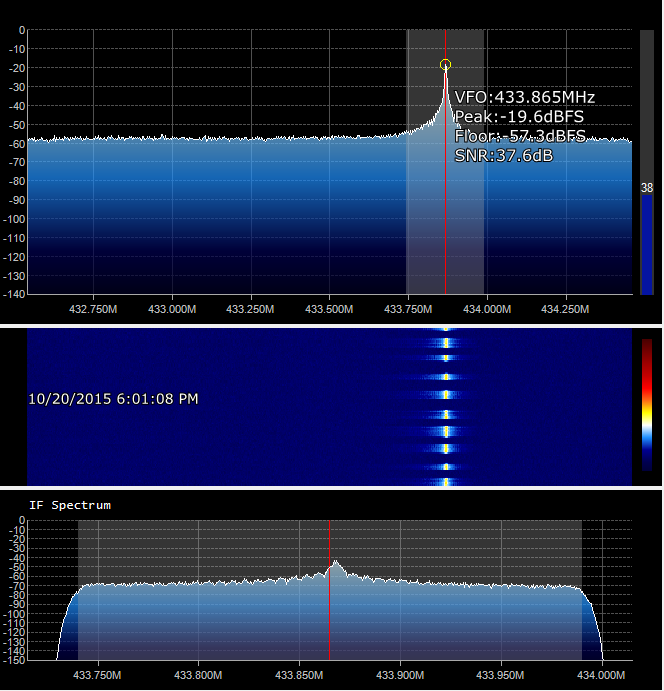}
\caption{ ISM 433 MHz device shown in Fig. \ref{fig:EtekcityDevice}, monitored using SDR\# software and captured using a HackRF receiver \cite{sdr4,hackrf}.}
\label{fig:433MhzSdrSharp}
\end{figure}

The SDR\# (SDRSharp) \cite{sdr4} is the popular software program that is used with the RTL-SDR. The software platform is free, fast and fairly easy to use. The 433 MHz ISM band signal captured is shown in Fig. \ref{fig:433MhzSdrSharp}. The sub-window shows the FFT observed, which is at 433.865 MHz with SNR of 37.6 dB. The signal captured is on-off keying (OOK) signal which is type of amplitude shift keying (ASK) modulation.

\subsection{OOK}
\label{ook}
\begin{figure} [!htbp]
\centering
\includegraphics[width=4cm,height=7cm,keepaspectratio]{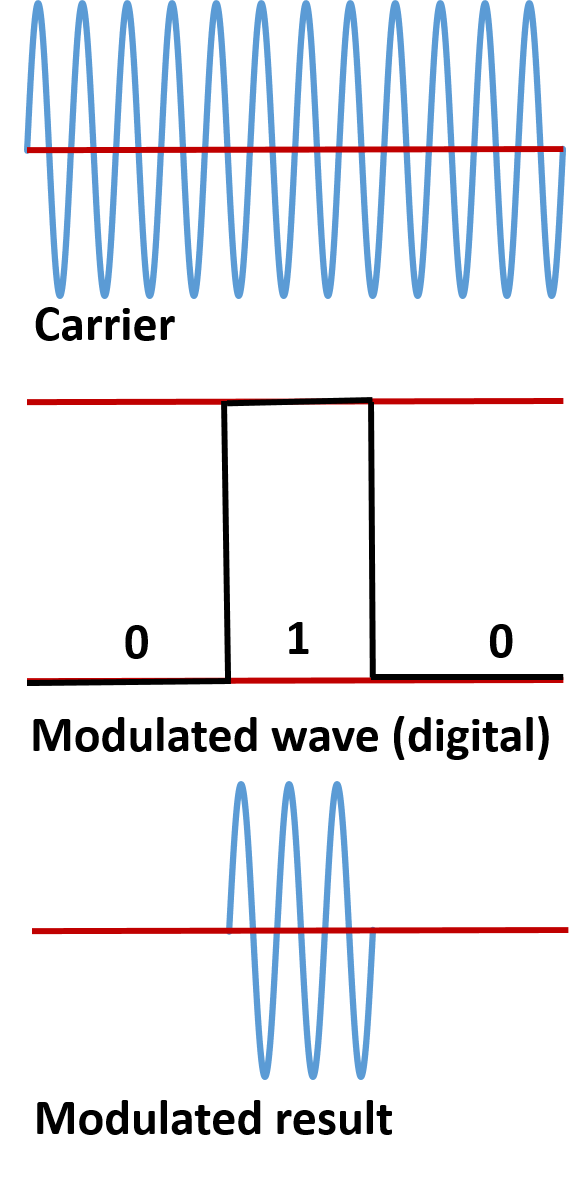}
\caption{OOK modulation}
\label{fig:ookSignal}
\end{figure}

OOK is the simplest form of ASK modulation that represents digital data. In its simplest form, the presence of a carrier for a specific duration represents a binary one, while its absence for the same duration represents a binary zero. Many of the sophisticated schemes vary these durations to convey additional information. It is analogous to unipolar encoding line code \cite{proakis2001digital}. Fig. \ref{fig:ookSignal}, shows the OOK modulation for a given digital signal.

OOK is also referred as continuous wave operation. For example, Morse code application can be transmitted using OOK encoding scheme. OOK has been used in the ISM bands to transfer data between device-to-device or device-to-computer. When compared to frequency-shift keying (FSK), OOK is more spectrally efficient but more sensitive to noise when using a regenerative receiver or a poorly implemented superheterodyne receiver \cite{gomez2010wireless}. When compared with binary phase shift keying (BPSK) signal, for a given data rate, the bandwidth of OOK signal and BPSK signal are equal. Furthermore, OOK is also used in optical communication systems.

\begin{figure} [!htbp]
\centering
\includegraphics[width=9cm,height=4cm,keepaspectratio]{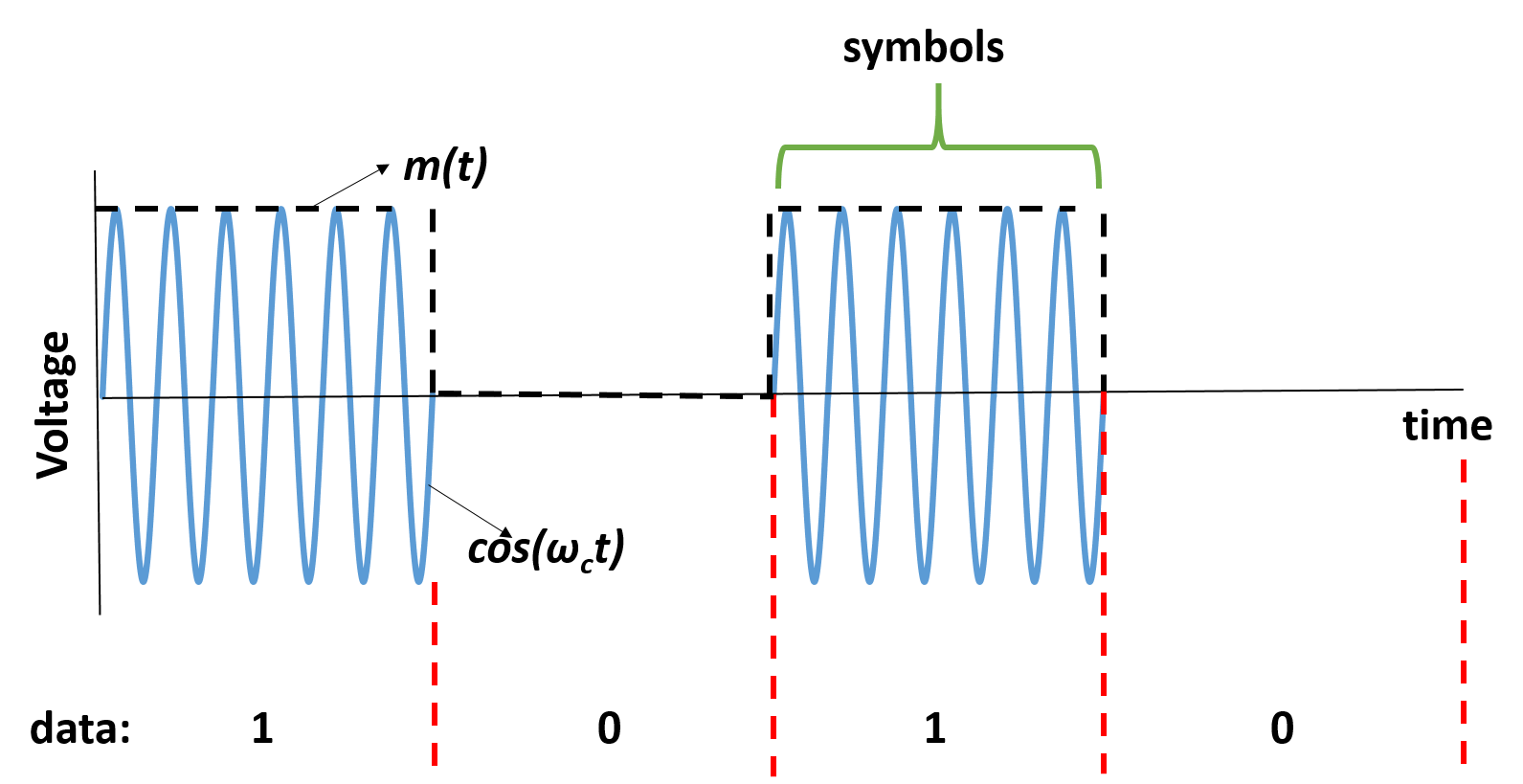}
\caption{OOK symbol mapping.}
\label{fig:ookSymbolMaping}
\end{figure}

\begin{equation}
\label{eq:ookSignal}
s(t) = A_cm(t)cos(\omega_ct)
\end{equation}

The OOK signal is represented by \ref{eq:ookSignal}, where $m(t)$ is unipolar baseband data signal and $A_c$ is the amplitude. The transmission bandwidth since OOK signal is twice the baseband bandwidth and AM-type signaling. If $r$ the raised cosine-rolloff filtering is used, the absolute bandwidth of the filtered binary signal is related to the bit rate $R$. The relationship between these parameters is given by \ref{eq:txBw}.

\begin{equation}
\label{eq:txBw}
B_T = (1+r)R
\end{equation}

\subsection{GNU radio implementation}
The GNU Radio is a free and open-source software development framework, which provides signal processing blocks to implement software radios on Linux machine. A GNU Radio application consists a set of signal processing blocks connected together that describe a data flow. This data flow is usually referred to as \textit{flowgraphs}. The reconfigurability of a system is a key feature of all the SDR system. The signal-processing software such as GNU Radio has the ability to provide general-purpose radio front-end, which handles the processing specific to the radio application. Hence, avoiding the usage of various radios designed for specific but distinct purposes.

\begin{figure} [!htbp]
\centering
\includegraphics[width=12cm,height=9cm,keepaspectratio]{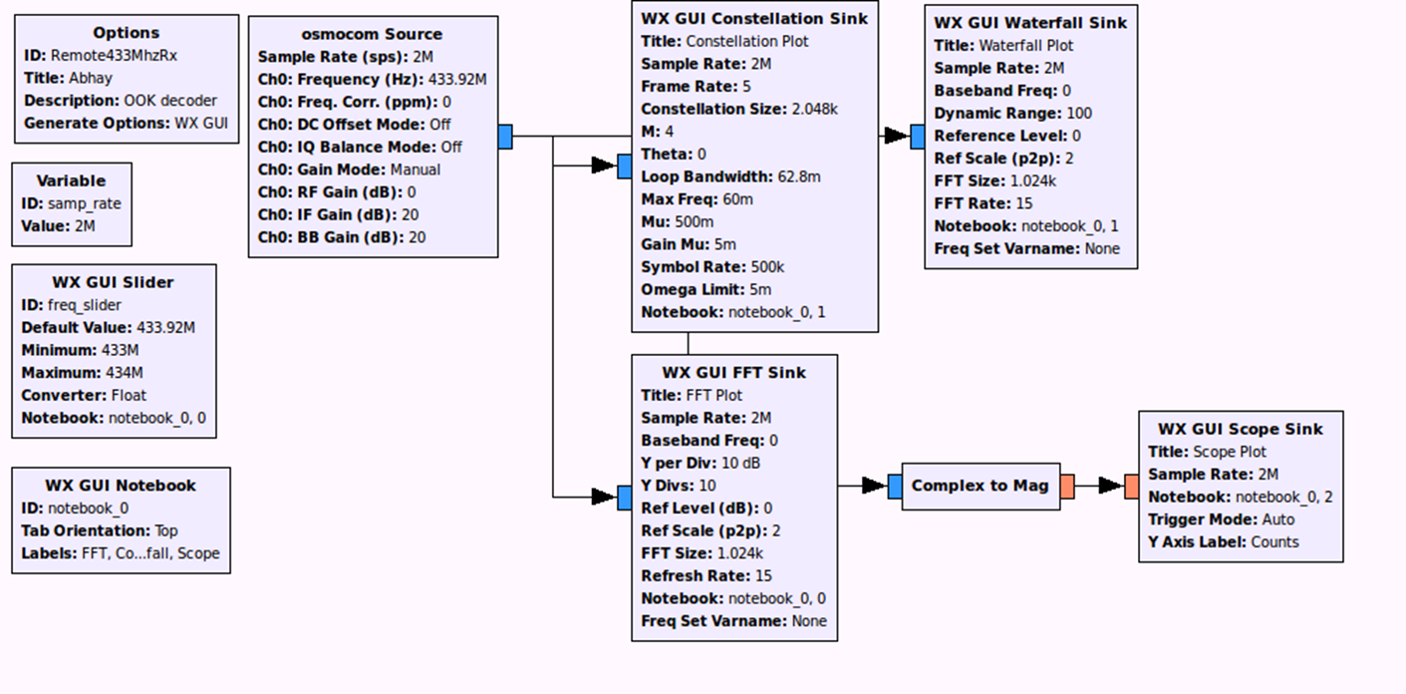}
\caption{OOK decoder flow diagram.}
\label{fig:ookDecoder}
\end{figure}

\begin{figure} [!htbp]
\centering
\includegraphics[width=8cm,height=7cm,keepaspectratio]{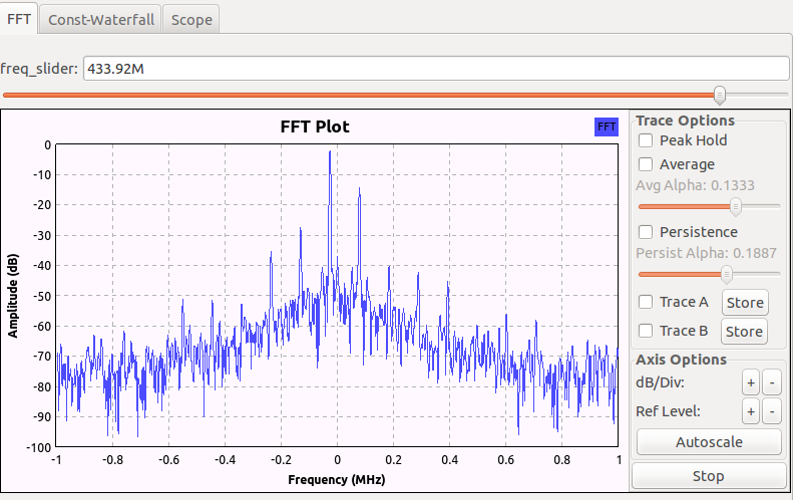}
\caption{FFT sink for captured OOK signal.}
\label{fig:fftSink}
\end{figure}

\begin{figure} [!htbp]
\centering
\includegraphics[width=8cm,height=7cm,keepaspectratio]{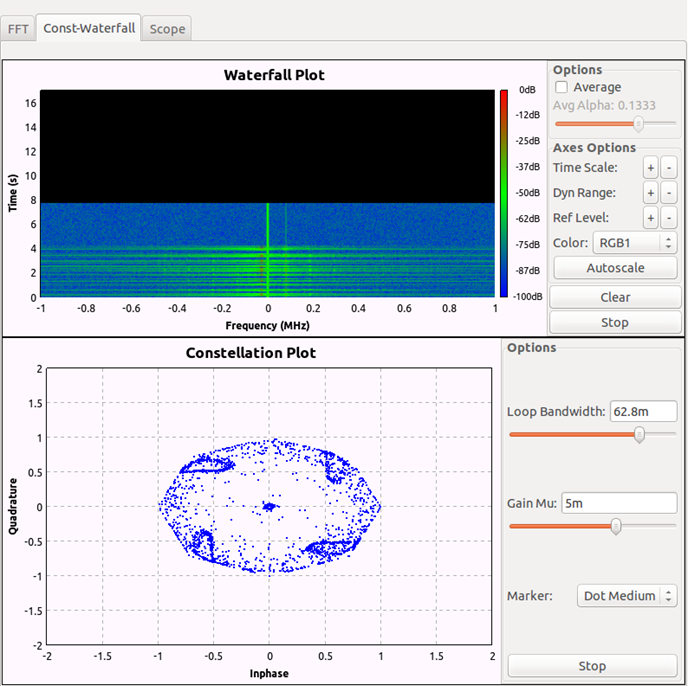}
\caption{Waterfall and constellation sink for captured OOK signal.}
\label{fig:waterfallSink}
\end{figure}

The Fig.~\ref{fig:ookDecoder} shows the OOK decoder flow diagram. The grc block for OOK decoder includes osmocom source for HackRF SDR for capturing of the signal. The source signals routed are to FFT sink, constellation sink, water sink and scope sink blocks. The FFT sink acts as a spectrum analyzer by doing a short time discrete fourier transform and the spectrum for the experimental setup \ref{fig:expSetup} is show in Fig.~\ref{fig:fftSink}. Furthermore, it can be also noticed that 433 MHz experimental spectrum observation using SDRSharp tool in Fig.~\ref{fig:433MhzSdrSharp} is in sync with FFT sink spectrum observation as seen in Fig.~\ref{fig:fftSink}. This implies the designed \textit{flowgraph} shown in Fig.~\ref{fig:ookDecoder} decodes OOK accurately. The other graphical sink used in this experimentation is waterfall sink shown in Fig.~\ref{fig:waterfallSink} which is consistent with the waterfall observation in Fig.~\ref{fig:433MhzSdrSharp}. The GNU Radio waterfall sink graphical representation of amplitude vs. frequency vs. time with amplitude represented as a variation in color.

\begin{figure} [!htbp]
\centering
\includegraphics[width=8.5cm,height=8cm,keepaspectratio]{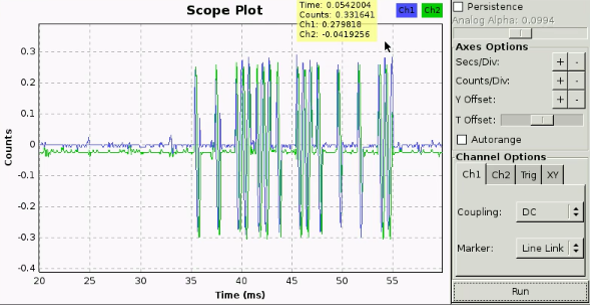}
\caption{OOK decoded signal in complex form.}
\label{fig:scopePlotComplexSig}
\end{figure}

\begin{figure} [!htbp]
\centering
\includegraphics[width=8.5cm,height=8cm,keepaspectratio]{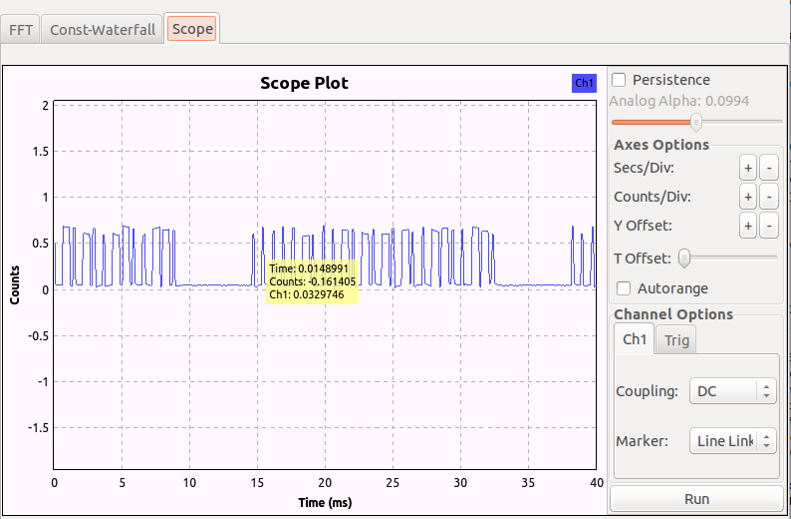}
\caption{OOK decoded signal in float/real form.}
\label{fig:scopePlotRealSig}
\end{figure}

\subsection{Result discussion}
By adding various sinks slowly-changing signals can be monitored. The scope sink in \textit{flowgraph}, can be modified to be display the decode OOK signal either in complex form or real/float form as shown in Fig.~\ref{fig:scopePlotComplexSig} and Fig.~\ref{fig:scopePlotRealSig}, respectively.  

The decoded signal observed on complex sink scope is in form burst of symbols. The burst of symbols are either short or long. Short burst is indicated by $0$ and long burst using $1$ binary symbol as described in one of the tutorials in \cite{hackrf}. Furthermore, the complex signal converted into real signals is as shown in Fig.~\ref{fig:scopePlotRealSig}. The signal observed in Fig.~\ref{fig:scopePlotRealSig} shows the OOK binary symbols that was transmitted by the wireless remote to control the electric switch. Based on the experiment carried out in this article, ON symbol burst is  \textbf{\textit{0000010001010101001100110}} and OFF symbol burst is \textbf{\textit{0000010001010101001111000}}. The control sequence varies from device to device. For instance, the example shown in one of the tutorials in \cite{hackrf} is different from what is been observed our current SDR experimentation. This concludes the control signal analysis i.e., ON-OFF signals from the Eteckcity device using SDR.  
        
\section{Future work}
\label{futureWork}
This paper has a limited scope and restricted to an overview of various ISM frequency bands. The goal in the upcoming survey paper would be to draw comparison between various wireless technologies in these ISM frequency bands. The future work as a letter article includes developing OOK encoder \textit{flowgraph} and send ON-OFF control signal using hackRF SDR to control electric switch.

\section{Conclusion}
\label{conclusion}
The ISM band is internationally reserved for the use of industrial, scientific and medical purposes. However, with other wireless communication protocols operating in the same frequency band, a harmonious outlook with existing devices is needed and this perspective is been worked upon by the scientific and engineering community. As discussed in this survey paper the ISM frequency bands are scattered and every ITU region having different regulatory policies, this call for unified policy-making for all the ISM frequency bands and the device operating in these bands. The technological advancements in the field of wireless communication has made it possible to share ISM band allocations with unlicensed and licensed operations.  

\bibliographystyle{IEEEtran}      
\bibliography{Citations}   

%
%

\end{document}